\documentclass[slac_one]{revtex4}
\usepackage{graphicx}
\usepackage{fancyhdr}
\begin{document}

\title{Physics of Neutrino Mass}

\author{Rabindra N. Mohapatra}
\affiliation{Department of Physics, University of Maryland, College Park,
MD-20742, USA}

\begin{abstract}
Recent discoveries in the field of neutrino oscillations have provided a
unique window into physics beyond the standard model. In this lecture, I
summarize how well we understand the various observations, what they
tell us about the nature of new physics and what we are likely to learn as
some of the planned experiments are carried out.
 \end{abstract}

\maketitle

\thispagestyle{fancy}

\section{INTRODUCTION}
For a long time, it was believed that neutrinos are massless, spin half
particles, making them drastically different from their other standard
model spin half cousins such as the charged leptons ($e, \mu,
\tau$) and the quarks ($u,d,s,c,t,b$), which are known to have mass. In
fact the masslessness of the neutrino was considered so sacred in
the 1950s and 1960s that the fundamental law of weak interaction
physics, the successful V-A theory for charged current weak processes was
considered to be intimately linked to this fact.

During the past decade, however, there have been a number of very exciting
observations involving neutrinos emitted in the process of solar burning,
produced during collision of cosmic rays in the atmosphere as well as
those produced in terrestrial sources such as reactors and accelerators
that have conclusively established that neutrinos not only have mass but
they also mix among themselves, like their counterparts $(e,\mu,\tau$) and
quarks, leading to the phenomenon of neutrino oscillations.
The detailed results of these experiments and their interpretation have
led to quantitative conclusions about the masses and the mixings, that
 have been discussed in other lectures\cite{kayser}. They have also been
summarized in
many recent reviews\cite{rev}. I will start with a brief summary of the
results.
I use the notation, where the flavor or
weak eigenstates are denoted by $\nu_{\alpha}$ (with $\alpha~=~e, {\mu},
{\tau},\cdot\cdot\cdot$), that are expressed in terms of the
mass eigenstates $\nu_{i}$ ($i=1, 2, 3,\cdot\cdot\cdot)$ as
follows: $\nu_{\alpha}
=~\sum_i
U_{\alpha i}\nu_i$. The $U_{\alpha i}$ can be also be expressed in terms
of mixing angles and phases as follows:
\begin{eqnarray}
 U~=~\pmatrix{c_{12}c_{13} & s_{12}c_{13} &
s_{13} e^{-i\delta} \cr
-s_{12}c_{23}-c_{12}s_{23}s_{13}e^{i\delta}
&c_{12}c_{23}-s_{12}s_{23}s_{13}e^{i\delta} & s_{23}c_{13}\cr
s_{12}s_{23}-c_{12}c_{23}s_{13}e^{i\delta}
&c_{12}s_{23}-c_{12}c_{23}s_{13}e^{i\delta} &
 c_{23}c_{13}}K
\end{eqnarray}
where $K~=~diag(1, e^{i\phi_1},e^{i\phi_2})$. This matrix characterizes
the weak charged current for leptons:
\begin{eqnarray}
{\cal L}_{wk}~=~\frac{g}{2\sqrt{2}}\bar{e}_\alpha U_{\alpha
i}\gamma_\mu (1+\gamma_5)\nu_i W^{\mu,-}~+~ h.c.
\end{eqnarray}
We denote the neutrino masses by $m_i$ ($i=1,2,3$).

\subsection{What we know about masses and mixings}
 Analysis of present
neutrino data tells us that (at the 3$\sigma$ level of confidence):
\begin{eqnarray}
sin^22\theta_{23}\geq 0.89\\ \nonumber
\Delta m^2_A\simeq 1.4\times 10^{-3}~ eV^2-3.3\times 10^{-3}~ eV^2\\
\nonumber
sin^2\theta_{12}\simeq 0.23-0.37\\ \nonumber
\Delta m^2_\odot\simeq 7.3\times 10^{-5}~ eV^2-9.1\times 10^{-5}~ eV^2\\
\nonumber
sin^2\theta_{13}\leq 0.047
\end{eqnarray}

While the mass differences that go into the discussion of oscillation
rate are fairly well determined (at least within the assumption of three
neutrinos and no exotic interactions), the situation with respect to
absolute values of masses is much less certain. There are three
possibilities:
\begin{itemize}

\item (i) Normal hierarchy i.e. $m_1\ll m_2 \ll m_3$. In this case,
we can deduce the value of $m_3 \simeq \sqrt{\Delta m^2_{23}}
\equiv \sqrt{\Delta m^2_A}\simeq 0.03-0.07$ eV. In this case $\Delta
m^2_{23}\equiv m^2_3-m^2_2 > 0$.
 The solar neutrino oscillation involves the two lighter levels. The mass
of the lightest neutrino is unconstrained. If $m_1\ll m_2$, then we get
the value of $m_2 \simeq \simeq 0.008$ eV.

\item (ii) Inverted hierarchy i.e. $m_1 \simeq m_2 \gg m_3$ with
$m_{1,2} \simeq \sqrt{\Delta m^2_{23}}\simeq 0.03-0.07$ eV. In this case,
solar neutrino oscillation takes place between the heavier levels and we
have $\Delta m^2_{23}\equiv m^2_3-m^2_2 < 0$.

\item (iii) Degenerate neutrinos i.e. $m_1\simeq m_2 \simeq m_3$.

\end{itemize}
The above conclusions do not depend on whether the neutrinos are Dirac or
Majorana fermions (Majorana fermions are their own anti-particles).

If neutrinos are Majorana fermions, they break lepton number by two units
and nuclear decay processes such as $(A,Z)\rightarrow (A, Z+2)+ e^- +e^-$
if allowed by kinematics can proceed. These are called $\beta\beta_{0\nu}$
process. The $\beta\beta_{0\nu}$ decay rate is directly
proportional to the neutrino mass since it is the neutrino mass term in
the Hamiltonian that breaks the lepton number symmetry. Present upper
limits on the $\beta\beta_{0\nu}$ decay rate puts an
 an upper limit on a particular combination of
masses and mixings (see the talk by G. Gratta at this
school\cite{gratta}):
\begin{eqnarray}
m_{eff}~=~\sum_i\left[U^2_{e i} m_i\right] \leq 0.3~eV
\end{eqnarray}
An important point here is that converting the neutrinoless double beta
upper limit to information about neutrino mass depends on the type of
spectrum\cite{vissani}. Fig. 1 gives the values of the effective
neutrino mass $m_{eff}$ predicted for the allowed range of mass
differences and mixings given by the present oscillation data.
It is clear that for the case of inverted hierarchy, one expects a lower
bound on $m_{eff}$ in the range 30 to 50 meV, whereas strictly speaking
for the normal hierarchy, this value can be zero due to CP violating
phases that can lead to possible cancellations.
 \begin{figure}[tbp]
\centerline{
\includegraphics[height=12.8cm,width=8cm]{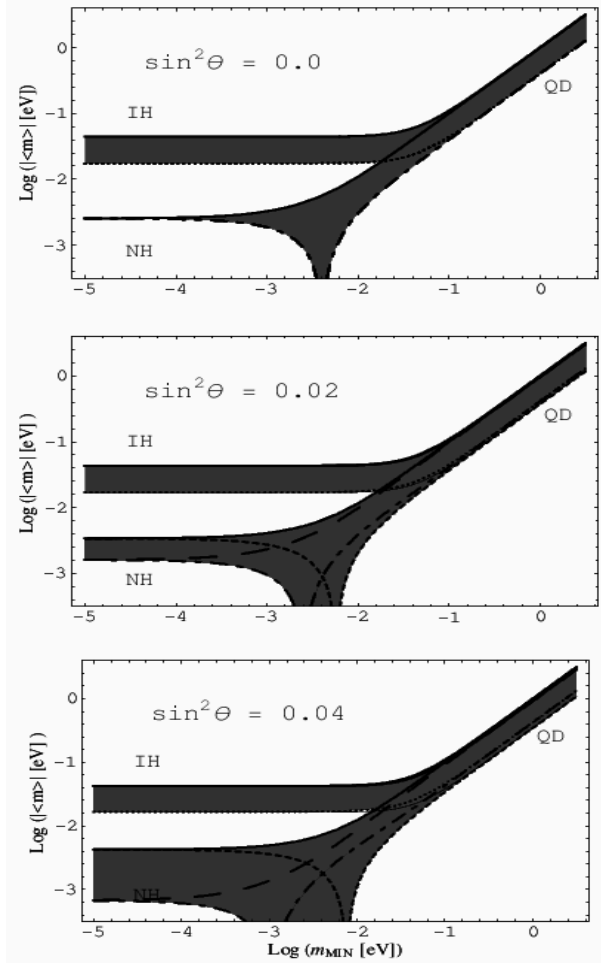}
}
\vspace{-3mm}
\caption{
The dependence of $ m_{\it eff}$ on $\langle
m\rangle_{min}$
in the case of the LMA-I solution,
for   normal and inverted hierarchy and
for the best fit values of the neutrino
oscillation parameters. Figure supplied by the authors of the last
reference in \cite{vissani}.}
\end{figure}
 At present there is also a claim of a positive signal for
$\beta\beta_{0\nu}$ decay at the level of few tenths of an eV that needs
to be confirmed\cite{klapdor}.
There are also limits from tritium beta decay end point search for
neutrino mass. In this case one gets a limit on the combination of
masses\cite{tri}:
\begin{eqnarray}
\sum_i|U_{e i}|^2 m^2_i \leq (2.2~eV)^2
\end{eqnarray}
From the above results, one can safely conclude that all known neutrinos
have masses in the eV to sub-eV range.

There are also limits from cosmological observations such as WMAP and SDSS
observations which put the limits in less than an eV range.
\begin{eqnarray}
\sum_i m_i \leq 0.4 ~eV
\end{eqnarray}

It is important to point out that a number of experiments are either
approved or planning or ongoing stage in the arena of tritium beta decay
(KATRIN), neutrinoless double beta decay (CUORE, MAJORANA, EXO
etc.)\cite{gratta} that will improve the above limits. From the domain of
cosmology, the PLANCK experiment will also tighten the upper limits on
neutrino masses.

 \subsection{Number of neutrinos}
In the above discussion we have assumed that there are only three neutrino
species. The question that arises is "How well do we know this ?". What we
know from laboratory experiments is that measurement of the Z-width at LEP
and SLC allows
only three species of light neutrinos that couple to the Z-boson. It is
however quite plausible to have additional neutrinos that are light and do
not couple
(or couple very, very weakly) to the Z-boson or the W boson. We will call
them sterile neutrinos
$\nu_s$ or $\nu'$. They are therefore unconstrained by the Z-width
data. However if
they mix with known neutrinos they can manifest themselves in the early
universe since the active neutrinos which are present in abundance in the
early universe can oscillate into the sterile neutrinos giving rise to a
density of $\nu_s$'s same as that of $\nu_{e,\mu,\tau}$. This will effect
the synthesis of Helium and Deuterium by enhancing the expansion rate of
the Universe. Thus our knowledge of the primordial Helium and Deuterium
abundance will then provide constraints on the total number of neutrinos
(active and sterile). The limit on the number of sterile neutrinos from
BBN depends on several inputs: the baryon to photon ratio $\eta\equiv
\frac{n_B}{n_\gamma}$ and the value of the He$^4$ fraction $Y_p$. The
first (i.e. $\eta$) is now very well determined by the WMAP observation of
the angular power spectrum\cite{WMAP}. The He$^4$ fraction $Y_p$ has
however been uncertain.

There have been new developments in this field. This has to do with our
knowledge of primordial Helium abundance, which is derived from the
analysis of low metallicity HII regions. It is now
believed\cite{olive} that there are more systematic  uncertainties in the
estimates of Helium abundance from these analyses than was previously
thought. The latest conclusion about  the number of neutrinos
depends on which observations (He$^4$, D$^2$ or WMAP)
are taken into consideration. For example,  He$^4$, D$^2$ and
$\eta_{CMB}$ together seem to give\cite{olive} $N_\nu \leq 4.44$
(compared to 3.3 before). One can therefore allow more than one
sterile neutrino mixing with the active neutrinos without conflicting
with cosmological observations. This has important implications
for interpretation of the positive neutrino oscillation signals observed
in the LSND experiment and now being tested by the Mini Boone experiment.
We discuss this at a later section of this paper.

\section{NEUTRINO MASS: DIRAC VRS MAJORANA}
In this section, I give a brief explanation of how to understand a
Majorana neutrino.
Let us write down the Dirac equation for an electron\cite{kayser}:
\begin{eqnarray}
i\gamma^{\lambda}\partial_{\lambda}\psi - m\psi =0
\end{eqnarray}
This equation follows from a free Lagrangian
\begin{eqnarray} {\cal L} =
i\bar{\psi}\gamma^{\lambda}\partial_{\lambda}\psi
-m\bar{\psi}\psi
\end{eqnarray}
The second term in the Lagrangian is the mass of the electron. However,
Lorentz invariance allows another bilinear for fermions that could also
act as a mass term i.e. $\psi^TC^{-1}\psi$, where $C$ is the charge
conjugation matrix. The difference between these two mass terms is that
the first one is invariant under a transformation of the form
$\psi\rightarrow e^{i\alpha}\psi$, whereas the second one is not.
To discriminate between the two kinds of mass terms, we need to know the
meaning of such a transformation: invariance under a phase transformation
implies the existence of a charge which is conserved (e.g. the elctric
charge, baryonic charge, leptonic charge etc.). Thus the presence of the
second kind of mass term means the theory breaks all symmetries. Further
note that if $\psi$ satisfies
the condition of being self charge conjugate, i.e. \begin{eqnarray}
 \psi = \psi^c \equiv C \bar{\psi}^T,
\end{eqnarray}
then the mass term $\bar{\psi}\psi$ reduces to the mass term
$\psi^TC^{-1}\psi$. Thus, the second mass term really implies that the
neutrinos are their own anti-particles. Furthermore, this
constraint reduces the number of independent components of the spinor
by a factor of two, since the particle and the antiparticle are now the
same particle. This mass term is called the Majorana mass in
contrast to the form $\bar{\psi}\psi$ which will be called Dirac
mass term.

Thus given a number of arbitrary spinors describing spin 1/2 particles,
one can write either only Dirac type mass terms or Majorana type mass
terms or both. Note that when a particle has a conserved quantum number
(e.g. electric charge for the electron), one cannot write a Majorana mass
term since it will break electric charge conservation. However for
particles such as the neutrino which are electrically neutral, both mass
terms are allowed in a theory. In fact one can stretch this argument
even further to say that if for an electrically neutral particle, the
Majorana mass term is not included, there must be an extra symmetry
in the theory to guarantee that it does not get generated in higher
orders. In general therefore, one would expect the neutrinos to be
Majorana fermions. That is what most extensions of the standard model
seem to predict. For a detailed discussion of this see \cite{rnm}.

At the moment we do not know if neutrinos are Dirac or Majorana fermions.
A crucial experiment that will determine this is the neutrinoless double
beta decay experiment\cite{gratta}. A positive signal in this experiment
conclusively establishes that neutrinos are Majorana fermions\cite{valle},
although contrary to popular belief, it will not be easy without further
experiments to determine the mass of the neutrino. The main reason for
this is that there could be heavy beyond the standard model particles that
could lead to $\beta\beta_{0\nu}$ decay withour at the same time giving a
``large'' enough neutrino mass\cite{rnm}.

An interesting question is: can we ever tell whether the neutrino is
a Dirac fermion ? One can of course never say whether a very tiny Majorana
mass term is present in the neutrino mass. This is in fact true for all
symmetries in Nature that we assume are exact e.g.  Lorentz
invariance, electric charge conservation etc. What we can however say is
whether the Dirac mass term dominates over the Majorana mass term
overwhelmingly.
 This can be done by a combination of the three
experiments: (i) $\beta\beta_{0\nu}$ decay experiments which are supposed
to reach the level of sensitivity of 30-50 milli eV, (ii) tritium beta
decay experiment KATRIN which is expected to push down the mass limit to
the level of 0.2 eV and (iii) a long
base line experiment that can presumably determine the sign of the
atmospheric mass
difference square. In the Table I we give the situations when one can
conclude that the neutrino is a Dirac particle\cite{njp} and when not.

\newpage

\begin{center}
{\bf Table I}
\end{center}

\begin{center}
\begin{tabular}{|c||c||c||c|}
\hline $\beta\beta_{0\nu}$ & $\Delta m^2_{23}$ & KATRIN &
Conclusion \\ \hline
yes & $>0$ & yes & Degenerate, Majorana \\
yes & $>0$ & No & Degenerate, Majorana\\
 & & & or normal or heavy exchange\\
yes & $<0$ & no & Inverted, Majorana \\
yes & $<0$ & yes & Degenerate, Majorana\\
no & $>0$ & no & Normal, Dirac or Majorana\\
no & $<0$ & no & Dirac\\
no & $<0$ & yes & Dirac \\
no & $>0$ & yes & Dirac \\ \hline
\end{tabular}
\end{center}
\noindent{{\bf Table Caption:} Conditions under which one can determine
when neutrino is a Dirac particle. Normal, inverted and degenerate refer
to the various mass patterns already discussed. }

Before closing this section, let us again summarize the open questions
raised by present data which need to be
addressed by future experiments:
\begin{quote}
$\bullet$ Are neutrinos Dirac or Majorana?\\
$\bullet$ What is the absolute mass scale of neutrinos?\\
$\bullet$ How small is $\theta_{13}$?\\
$\bullet$ How ``maximal'' is $\theta_{23}$?\\
$\bullet$ Is there CP Violation in the neutrino sector?\\
$\bullet$ Is the mass hierarchy inverted or normal?\\
$\bullet$ Is the LSND evidence for oscillation true?  Are there sterile
      neutrino(s)?
\end{quote}

It is important to emphasize that if the full menu of experiments being
proposed currently such as searches for neutrinoless double beta decay,
searches for $\theta_{13}$, precision measurements of $\theta_{12}$ and
$\theta_{23}$ using reactor and long baseline experiments, all these
questions would recieve answers.

\section{IMPLICATIONS FOR PHYSICS BEYOND THE STANDARD MODEL:}
These discoveries involving neutrinos, which have provided the
first evidence for physics beyond the standard model, have raised a
number of challenges for theoretical physics.
Foremost among them are, (i) an  understanding of the smallness of
neutrino masses and (ii) understanding the vastly different pattern of
mixings among neutrinos from the quarks. Specifically, a key question
is whether it is possible
to reconcile the large neutrino mixings with small quark mixings in grand
unified frameworks suggested by supersymmetric gauge coupling
unifications that unify quarks and leptons.

\subsection{  Why neutrino mass requires physics beyond
the standard model ?}

We will now show that in the standard model, the neutrino mass vanishes
to all orders in perturbation theory as well as nonperturbatively. The
standard model is based on the gauge group
$SU(3)_c\times SU(2)_L\times U(1)_Y$ group under which the quarks and
leptons transform as described in the Table II.

\newpage

\begin{center}
{\bf Table II}
\end{center}

\begin{center}
\begin{tabular}{|c||c|}
\hline\hline
 Field &  gauge  transformation \\ \hline\hline
 Quarks $Q_L$ & $(3,2, {1\over 3})$\\
 Righthanded up quarks $u_R$ &  $(3, 1, {4\over 3})$ \\
Righthanded down quarks  $ d_R$ &  $(3, 1,-\frac{2}{3})$\\
Lefthanded  Leptons $L$ & $(1, 2 -1)$ \\
 Righthanded leptons  $e_R$ & $(1,1,-2)$ \\
Higgs Boson $\bf H$ & $(1, 2, +1)$ \\
Color Gauge Fields  $G_a$ & $(8, 1, 0)$ \\
Weak Gauge Fields  $W^{\pm}$, $Z$, $\gamma$ & $(1,3+1,0)$ \\
\hline\hline
\end{tabular}
\end{center}

\noindent {\bf Table caption:} The assignment of particles to the standard
model gauge group $SU(3)_c\times SU(2)_L\times U(1)_Y$.
The electroweak symmetry $SU(2)_L\times U(1)_Y$ is broken by the vacuum
expectation of the Higgs doublet $<H^0>=v_{wk}\simeq 246$ GeV, which gives
mass to the gauge bosons and the fermions, all fermions except the
neutrino. Thus the neutrino is massless in the standard model, at the tree
level.
 There are several questions that arise at this stage. What happens
when one goes beyond the above simple tree level approximation ? Secondly,
do nonperturbative effects change this tree level result ? Finally, how to
judge how this result will be modified when the quantum gravity effects
are included ?

The first and second questions are easily answered by using the B-L
symmetry of the standard model. The point is that since the standard model
has no $SU(2)_L$ singlet neutrino-like field, the only possible mass terms
that are allowed by Lorentz invariance are of the form
$\nu^T_{iL}C^{-1}\nu_{jL}$, where $i,j$ stand for the generation index and
$C$ is the Lorentz charge conjugation matrix. Since the $\nu_{iL}$ is part
of the $SU(2)_L$ doublet field and has lepton number +1, the above
neutrino mass term transforms as an $SU(2)_L$ triplet and furthermore, it
violates total lepton number (defined as $L\equiv L_e+L_{\mu}+L_{\tau}$)
by two units. However, a quick look at the standard model Lagrangian
convinces one that the model has exact lepton number symmetry after
symmetry breaking; therefore such terms can never arise in perturbation
theory.
Thus to all orders in perturbation theory, the neutrinos are massless.
As far as the nonperturbative effects go, the only known source is the
weak instanton effects. Such effects could effect the result if they
broke the lepton number symmetry. One way to see if such breaking
weak instanton effects. Such effects could effect the result if they
broke the lepton number symmetry. One way to see if such breaking
occurs is to look for anomalies in lepton number current conservation from
triangle diagrams. Indeed $\partial_{\mu}j^{\mu}_{\ell}= c W \tilde{W} +
c' B\tilde{B}$ due to the contribution of the leptons to the triangle
involving the lepton number current and $W$'s or $B$'s. Luckily, it turns
out that the anomaly contribution to the baryon number current
nonconservation has also an identical form, so that the $B-L$ current
$j^{\mu}_{B-L}$ is conserved to all orders in the gauge couplings. As a
consequence, nonperturbative effects from the gauge sector cannot induce
$B-L$ violation. Since the neutrino mass operator described above violates
also $B-L$, this proves that neutrino masses remain zero even in the
presence of nonperturbative effects.

Let us now turn to the effect of gravity. Clearly as long as we treat
gravity in perturbation theory, the above symmetry arguments hold since
all gravity coupling respect $B-L$ symmetry. However, once nonperturbative
gravitational effects e.g black holes and worm holes are
included, there is no guarantee that global symmetries will
be respected in the low energy theory. The intuitive way to appreciate the
argument is to note that throwing baryons into a black hole does not lead
to any detectable consequence except thru a net change in the baryon
number of the universe. Since one can throw in an arbitrary numnber of
baryons into the black hole, an arbitrary information loss about the net
number of missing baryons would prevent us from defining a baryon
number of the visible
universe- thus baryon number in the presence of a black hole can not be an
exact symmetry. Similar arguments can be made for any global charge such
as lepton number in the standard model. A field theoretic parameterization
of this statement is that the effective low energy Lagrangian for the
standard model in the presence of black holes and worm holes etc must
contain baryon and lepton number violating terms. In the context of the
standard model, the only such terms that one can construct are
nonrenormalizable terms of the form $~LH LH/M_{P\ell}$. After gauge
symmetry breaking, they lead to neutrino masses; however these masses are
at most of order $~v^2_{wk}/M_{P\ell}\simeq 10^{-5}$ eV.
But
as we discussed in the previous section, in order to solve the atmospheric
neutrino problem, one needs masses at least three orders of magnitude
higher.

Thus one must seek physics beyond the standard model to explain observed
evidences for neutrino masses. While there are many possibilities that
lead to small neutrino masses of both Majorana as well as Dirac kind, here
we focus on the possibility that there is a heavy right handed
neutrino (or neutrinos) that  lead to a small
neutrino mass. The resulting
mechanism is known as the seesaw mechanism \cite{seesaw1} and leads to
neutrino being a Majorana particle.

\subsection{Seesaw mechanism}
The basic idea of seesaw mechanism is to have a  minimal extension of the
standard model that add one heavy right handed neutrino per family.
 In this case $\nu_L$ and $\nu_R$ can
form a mass term; but apriori, this mass term is like the mass terms for
charged leptons or quark masses and will therefore involve the weak scale.
If we call the corresponding Yukawa coupling to be $Y_\nu$, then the
neutrino mass is $m_D=Y_\nu v/\sqrt{2}$. For a neutrino mass in the eV
range
requires that $Y_\nu \simeq 10^{-11}$ or less. Introduction of such small
coupling constants into a theory is generally considered unnatural and a
sound theory must find
a symmetry reason for such smallness. As already already
alluded to before, seesaw mechanism\cite{seesaw1}, where we introduce a
singlet Majorana mass term for the right handed neutrino is one way to
achieve this goal. What we have in this case is a
$(\nu_L,\nu_R)$ mass matrix which has the form:
\begin{eqnarray}
M=\left(\begin{array}{cc} 0 & M_\nu^D \\
M^{T,D}_\nu & M_R\end{array}\right)
\end{eqnarray}
 The light neutrino mass matrix obtained by
integrating out heavy right-handed neutrinos is given by
\begin{equation}
{M}_\nu = - M_{\nu}^D M_R^{-1} (M_\nu^D)^T,
\end{equation}
where $M_\nu^D$ is the Dirac neutrino mass matrix and $M_R$ is the
right-handed Majorana mass matrix.
Since $M_R$ is not constrained by the standard model symmetries, it is
natural to choose it to be at a scale much higher than the weak scale,
leading to a small mass for the neutrino.
This provides a natural way to understand a small neutrino mass without
any unnatural adjustment of parameters of a theory.
A question that now arises is: what is the meaning of the new scale $M_R$
?

\section{PHYSICS OF THE SEESAW MECHANISM}
Inclusion of the right handed neutrino to the standard model open up a
whole new way of looking at physics beyond the standard model and
transforms the dynamics of the standard model in a profound
way. To clarify what we mean, note that in the standard model (that does
not contain a $\nu_R$) the $B-L$ symmetry is only linearly anomaly free
i.e. $Tr[(B-L)Q^2_a]=0$ where $Q_a$ are the gauge generators of the
standard model but $Tr(B-L)^3\neq 0$. This means that $B-L$ is only a
global symmetry and cannot be gauged. However as soon as the $\nu_R$ is
added to the standard model, one gets $Tr[(B-L)^3]=0$ implying that the
B-L symmetry is now gaugeable and one could choose the gauge group of
nature to be either $SU(2)_L\times U(1)_{I_{3R}}\times U(1)_{B-L}$ or
$SU(2)_L\times SU(2)_R\times U(1)_{B-L}$, the latter being the gauge group
of the left-right symmetric models\cite{moh}. Furthermore the presence of
the $\nu_R$ makes the model quark lepton symmetric and leads to a
Gell-Mann-Nishijima like formula for the elctric charges\cite{marshak}
i.e.
\begin{eqnarray}
Q= I_{3L}+I_{3R}+\frac{B-L}{2}
\end{eqnarray}
The advantage of this formula over the charge formula in the standard
model charge formula is that in this case all entries have a physical
meaning. Furthermore, it leads naturally to Majorana nature of neutrinos
as can be seen by looking at the distance scale where the $SU(2)_L\times
U(1)_Y$ symmetry is valid but the left-right gauge group is broken. In
that case, one gets
\begin{eqnarray}
\Delta Q=0= \Delta I_{3L}:\\ \nonumber
\Delta I_{3R}~=~-\Delta \frac{B-L}{2}
\end{eqnarray}
We see that if the Higgs fields that break the left-right gauge group
carry righthanded isospin of one, one must have $|\Delta L| = 2$ which
means that the neutrino mass must be Majorana type and the theory will
break lepton number by two units. As we see below this Majorana mass
arises via the seesaw mechanism as was first shwon in the last reference
in \cite{seesaw1}. It also further connects the nonzero neutrino mass to
the maximal V-A character of the weak interaction forces. To show this, we
discuss the left-right models and show how neutrino small neutrino mass
arises in this model via the seesaw mechanism and how it is connected to
the scale of parity violation. This may provide one answer to the raised
 in C. Quigg's lecture at this institute regarding why weak interactions
are maximally parity violating unlike any other force in Nature.

\subsection{Left-right symmetry, neutrino mass and origin of V-A weak
interactions}

The left-right symmetric theory is basrd on the gauge group
SU$(2)_L \, \times$ SU$(2)_R \, \times$ U$(1)_{B-L}$ with quarks and
leptons transforming as doublets under SU$(2)_{L,R}$.
In Table III, we denote the quark, lepton and Higgs
fields in the theory along with their transformation properties
under the gauge group.
~~~~~~~~~~
\begin{center}
{\bf Table III}
\end{center}

\begin{center}
\begin{tabular}{|c|c|} \hline\hline
Fields           & SU$(2)_L \, \times$ SU$(2)_R \, \times$ U$(1)_{B-L}$ \\
                 & representation \\ \hline
$Q_L$                & (2,1,$+ {1 \over 3}$) \\
$Q_R$            & (1,2,$ {1 \over 3}$) \\
$L_L$                & (2,1,$- 1$) \\
$L_R$            & (1,2,$- 1$) \\
$\phi$     & (2,2,0) \\
$\Delta_L$         & (3,1,+ 2) \\
$\Delta_R$       & (1,3,+ 2) \\
\hline\hline
\end{tabular}
\end{center}

\noindent{\bf Table caption} Assignment of the fermion and Higgs
fields to the representation of the left-right symmetry group.

\bigskip

The first task is to specify how the left-right symmetry group breaks to
the standard model i.e. how one
breaks the $SU(2)_R\times U(1)_{B-L}$ symmetry so that the successes of
the standard model
including the observed predominant V-A structure of weak interactions at
low energies is reproduced. Another question of naturalness that also
arises simultaneously is that since the charged fermions and the
neutrinos are treated completely symmetrically (quark-lepton symmetry)
in this model, how does one understand the smallness of the neutrino
masses compared to the other fermion masses.

It turns out that both the above problems of the LR model have a common
solution. The process of spontaneous breaking of the $SU(2)_R$ symmetry
that suppresses the V+A
currents at low energies also solves the problem of ultralight neutrino
masses. To see this let us write the Higgs fields explicitly:
\begin{eqnarray}
\Delta~=~\left(\begin{array}{cc} \Delta^+/\sqrt{2} & \Delta^{++}\\
\Delta^0 & -\Delta^+/\sqrt{2} \end{array}\right); ~~
\phi~=~\left(\begin{array}{cc} \phi^0_1 & \phi^+_2\\
\phi^-_1 & \phi^0_2 \end{array}\right)
\end{eqnarray}
 All these
Higgs fields have Yukawa couplings to the fermions given symbolically as
below.
\begin{eqnarray}
{\cal L_Y}= h_1 \bar{L}_L\phi L_R +h_2\bar{L}_L\tilde{\phi}L_R\nonumber \\
+ h'_1\bar{Q}_L\phi Q_R +h_2'\bar{Q}_L\tilde{\phi}Q_R
\nonumber\\
+f(L_LL_L\Delta_L +L_RL_R\Delta_R) +~ h.c. \end{eqnarray}
The $SU(2)_R\times U(1)_{B-L}$ is broken down to the standard model
hypercharge $U(1)_Y$ by choosing $<\Delta^0_R>=v_R\neq 0$ since this
carries
both $SU(2)_R$ and $U(1)_{B-L}$ quantum numbers. It gives mass to the
charged and neutral righthanded gauge bosons i.e. $M_{W_R}= gv_R$ and
$M_{Z'}=\sqrt{2} gv_R cos\theta_W/\sqrt{cos 2\theta_W}$. Thus by
adjusting the value of $v_R$ one can suppress the right handed current
effects in both neutral and charged current interactions arbitrarily
leading to an effective near maximal left-handed form for the charged
current weak interactions.

The fact that at the same time the neutrino masses also become small can
be
seen by looking at the form of the Yukawa couplings. Note that the f-term
leads to a mass for the right handed neutrinos only at the scale $v_R$.
Next as we break the standard model symmetry by turning on the vev's for
the $\phi$ fields as $Diag<\phi>=(\kappa, \kappa')$, we not only
give masses to the $W_L$ and the $Z$ bosons but also to the quarks and the
leptons. In the neutrino sector the above Yukawa couplings after
$SU(2)_L$ breaking by $<\phi>\neq 0$ lead to the so called Dirac masses
forthe neutrino
connecting the left and right handed neutrinos. In the two component
neutrino language, this leads to the following mass matrix for the
$\nu, N$ (where we have denoted the left handed neutrino by $\nu$ and the
right handed component by $N$).
\begin{eqnarray}
M=\left(\begin{array}{cc} 0 & h\kappa \\
h\kappa & fv_R\end{array}\right)
\end{eqnarray}
Note that $m_D$ in previous discussions of the seesaw formula (see Eq. ())
is given by $m_D=h\kappa$, which links it to the weak scale and the mass
of the RH neutrinos is given by
$M_R=f v_R$, which is linked to the local B-L symmetry. This
justifies keeping RH neutrino mass at a scale lower than the Planck mass.
It is therefore fair to assume that seesaw mechanism coupled with
observations of neutrino oscillations are a strong indication of the
existence of a local B-L symmetry far below the Planck scale.

\subsection{Parity symmetry and type II seesaw}
In deriving the above seesaw formula for neutrino masses, it has been
assumed that the vev of the lefthanded triplet is zero so that the
$\nu_L\nu_L$ entry of the neutrino mass matrix is zero. However, in the
left-right model which provide an explicit
derivation of this formula, there is an induced ve for the $\Delta^0_L$
of order $<\Delta^0_L> = v_T\simeq \frac{v^2_{wk}}{v_R}$. In the
left-right models, this this arises from the presence of a coupling in the
Higgs potential of the form
$\Delta_L\phi\Delta^{\dagger}_R\phi^{\dagger}$. In the  presence
of the $\Delta_L$ vev, the seesaw formula undergoes a fundamental
change. One can have two types of seesaw formulae depending on whether
the $\Delta_L$ has vev or not. The new seesaw
formula now becomes:
\begin{equation}
{ M}_\nu^{\rm II} = M_L - M_\nu^D M_R^{-1} (M_\nu^D)^T,
\end{equation}
where $M_L = f v_L$ and $M_R=f v_R$, where $v_{L,R}$ are the vacuum
expectation values of Higgs fields that couple to the right and lefthanded
neutrinos.
This formula for the neutrino mass matrix is called type II seesaw formula
\cite{seesaw2}. In Fig. 2, we give the diagrams that in a parity symmetric
theory lead to the type II seesaw formula.

\begin{figure}[tbp]
  \centering
  \includegraphics[width=\textwidth]{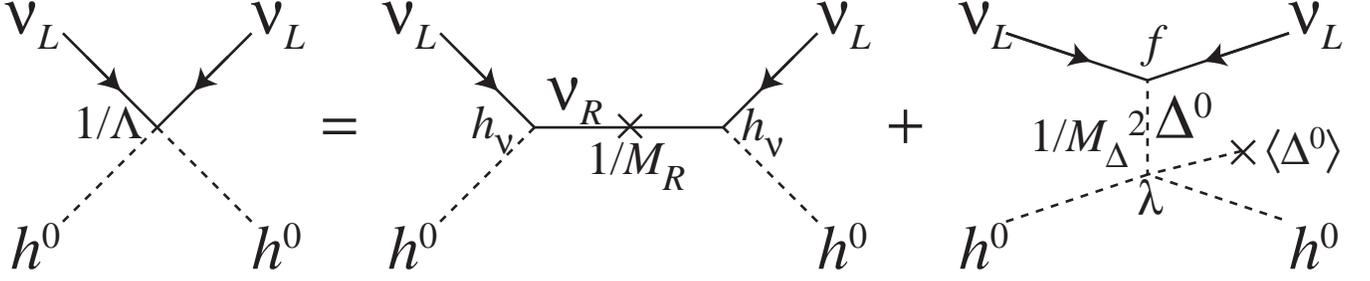}
  \caption{Seesaw mechanism: the first diagram involves the exchange of
the heavy right handed neutrino and by itself leads to type I seesaw
whereas the second figure gives the extra contribution to seesaw formula
in parity symmetric theories and leads to type II seesaw. Of the two
$\Delta$ fields in this figure, the one that has a vev is the $\Delta_R$
field of Table III and the one that connects the two vertices is the
$\Delta_L$ of Table III. }
  \label{fig:seesaw}
\end{figure}

One may perhaps get some hint as to which type of seesaw formula is valid
in Nature once the neutrino spectrum is determined.
In the type I seesaw formula, what appears is the square of the
Dirac neutrino mass matrix which in general expected to have the same
hierarchical structure as the corresponding charged fermion mass matrix.
In fact in some specific GUT models such as SO(10), $M_D=M_u$ which
validates this conjecture leading to
 the common statement that neutrino masses given by the seesaw
formula are hierarchical
i.e.
$m_{\nu_e}\ll m_{\nu_{\mu}}\ll m_{\nu_{\tau}}$ and even a more model
dependent statement that $m_{\nu_e} : m_{\nu_{\mu}} : m_{\nu_{\tau}}=
m^2_u : m^2_c : m^2_t$.

On the other hand in the type II seesaw formula, there is no
reason to expect a hierarchy and in fact if the neutrino masses turn out
to be degenerate (as discussed before as one possibility), one possible
way to understand this may be to use the type II seesaw formula.
Since the type II seesaw formula is a reflection of the parity
invariance of the theory at high energies, evidence for it would point
very strongly towards left-right symmetry at high energies. It also must
be stated that a hierarchical mass spectrum could result in either type of
seesaw formula. For an example of type II seesaw formula with hierarchical
spectrum, see the SO(10) model below.

The generic seesaw models lead to a number interesting phenomenological
and cosmological consequence that we will not discuss in this talk:

\begin{itemize}

\item Seesaw mechanism embedded into a supersymmetric framework with
supersymmetry broken at the weak scale leads to nonvanishing lepton flavor
violation. The detailed predictions for flavor violation depends on
specific assumptions. But still one would generally expect in these models
that the branching ratio for $\mu\rightarrow e+\gamma$ in these models is
expected to be above $10^{-14}$, which is the current goal of the PSI MEG
experiment.\cite{masiero}.

\item The decay of right handed neutrinos in conjunction with CP violation
in the right handed neutrino sector has been a very viable mechanism for
origin of matter via lkeptogenesis\cite{yana}.

\item The above hig scale CP violation could lead to measurable electric
dipole moments for leptons\cite{edm}.

\end{itemize}

\section{UNDERSTANDING LARGE MIXINGS}
While the seesaw formula provides an elegant way to understand the small
neutrino masses, it throws no light on the nature of the neutrino
mixings. The reason essentially is that for three active neutrinos, the
seesaw formula involves 18 unknown parameters whereas thye number of
observables for neutrinos is nine including all three phases. One must
therefore make specific assumptions or models in order to understand
mixings\cite{king}.

The neutrino mixing angles get contributions from the
mass matrices for the charged leptons as well as neutrinos. Since we can
choose
an arbitrary basis for either the charged leptons or the neutrinos without
effecting weak interactions, it is often convenient to work in a basis
where charged lepton mass matrix is diagonal.  A fundamental theory
can of course determine the structure of both the charged lepton and the
neutrino mass matrices and
therefore will lead to predictions about lepton mixings. However, in the
absence of such a theory, if one wants to adopt a model independent
approach and look for symmetries
that may explain say the maximal value of $\theta_{23}$ or large
$\theta_{12}$ etc., it is useful to work in a basis
where charged leptons are mass eigenstates
 and hope that any symmetries for leptons revealed in this
basis are true or approximate symmetries of Nature.

It could of course be that the large mixings are the result of some
dynamical mechanism e.g. radiative corrections or grand unification
and not a symmetry. In such a case, there is no need to start with a
particular basis. However, we must then find some
characxteristic experimental signatures that
could point towards such a theory.

In any case, it is necessary to look for signatures of the two approaches
to mixing angles i.e. whether it is the symmetry that is
responsible for large mixings or dynamics. Below, we describe, one give
several examples where either  a symmetry or some dynamical reason leads
to large mixings.

\subsection{$\mu-\tau$ symmetry and large atmospheric mixings}
 In the basis where charged leptons are mass eigenstates, a symmetry that
has proved useful in understanding maximal atmospheric neutrino mixing is
 $\mu\leftrightarrow \tau$ interchange symmetry\cite{mutau}. The mass
difference between the muon and the tau lepton of course breaks this
symmetry. So we expect this symmetry to be an approximate one. It may
however happen that the symmetry is truly exact at a very high scale; but
at low mass scales, the effective theory only has the $\mu-\tau$ symmetry
in the
neutrino couplings but not in the charged lepton sector so that we
have $m_\tau \gg m_\mu$\cite{grimus}.

To see how the symmetry of the mass matrix affects the mixing
matrix, let us consider the case of only two neutrino generations
i.e. that of $\mu$ and $\tau$. Experiments indicate that the
atmospheric mixing angle is very nearly maximal i.e. $\theta_A~=~
\pi/4$.  Working in the basis where the charged lepton mass
matrix is diagonal, it is obvious that the nautrino Majorana mass
matrix that gives maximal mixing is:
\begin{eqnarray}
{\cal M}^{(2)}_\nu~=~\pmatrix{a & b\cr b & a}.
\end{eqnarray}
This mass matrix has $\mu-\tau$ interchange symmetry. Smallness of solar
neutrino mass difference implies that we can write $b=-1$ and
$a=1+\epsilon$. Clearly, if such a symmetry is responsible for maximal
atmospheric mixing angle, it will be against the spirit of quark lepton
unification that is a fundamental part of the idea of grand unification.
Since there also grand unified models that can lead to near maximal
mixing, an important question is: how to distringuish a lepton specific
symmetry approach from a general quark-lepton unified GUT approach.

To answer this question let us extend the above symmetry discussion to the
case of three neutrinos. We then have
\begin{eqnarray}
{\cal M}_\nu~=~\frac{\sqrt{\Delta m^2_A}}{2}\pmatrix{c\epsilon
&d\epsilon &b\epsilon\cr d\epsilon & 1+a\epsilon & -1 \cr
b\epsilon & -1 & 1+\epsilon}
\end{eqnarray}
Note that if $a=1$ and $b=d$, this mass matrix has $\mu-\tau$
symmetry and leads to large solar mixing. It also predicts
$\theta_{13}=0$. However as (i) $a\neq 1$ or (ii) $b\neq d$, we get
nonzero $\theta_{13}$ and for case (ii) $\theta_{13}\sim \sqrt{\Delta
m^2_\odot/\Delta m^2_A}$ and $\theta_{13}\sim {\Delta
m^2_\odot/\Delta m^2_A}$  in case (i)\cite{theta13}.

In comparision, in a dynamical approach such as those based on grand
unified theories, we would have to have a mass matrix of type in
Eq. (19) but since there is no symmetry, we would expect both $a\neq 1$
and
$b\neq d$. So that we would expect  $\theta_{13}\geq \sqrt{\Delta
m^2_\odot/\Delta m^2_A}$. Since the next generation of neutrino
experiments are expected to push the limit on $\theta_{13}$ down to the
level of $0.04$ or so\cite{theta13e}, it should provide a hint as to
whether the GUT approach or the symmetry approach is more promising.

\subsection{ Inverted hierarchy and $L_e-L_\mu-L_\tau$ symmetry and large
solar mixing}
Another very natural way to understand large mixings is to assume
the symmetry $L_e-L_\mu-L_\tau$ for neutrinos. This symmetry as we see
below, leads to an
inverted mass hierarchy for neutrinos, which is therefore a clear
experimental prediction of this approach.
Consider the mass matrix
\begin{eqnarray}
{ M}_\nu=m_0~\left(\begin{array}{ccc} \epsilon &
c & s\\ c & \epsilon & \epsilon\\ s & \epsilon &
\epsilon\end{array}\right).
\end{eqnarray}
where $c=cos\theta_A$ and $s=sin\theta_A$. This mass matrix leads to
mixing angles that are completely consistent with all data. The mass
pattern in this case is inverted i. e. the two mass eigenstates
responsible for solar neutrino oscillation are nearly degenerate in mass
and the third neutrino mass is much smaller and could be zero. In the
limit of $\epsilon\rightarrow 0$, this mass matrix has $L_e-L_\mu-L_\tau$
symmetry. One therefore might hope that if inverted hierarchy structure is
confirmed, it may provide evidence for this leptonic symmetry which
can be an important clue to new physics beyond the standard model.
In  fact large departure of the solar mixing angle from its
 maximal value means that  $L_e-L_\mu-L_\tau$ symmetry must be badly
broken\cite{emutau}.

As in the above example, in the mass matrix in Eq. (), when we set
$cos\theta_A= sin\theta_A=\frac{1}{sqrt{2}}$, the theory becomes
$\mu-\tau$ symmetric and we get $\theta_{13}=0$. Therefore there is a
correlation between $\theta_A$ and $\theta_{13}$ in the case for this
case.

\subsection{Quark-lepton complementarity and large solar mixing}
There has been a recent suggestion\cite{raidal} that perhaps the large
but not maximal solar mixing angle is related to physics of the quark
sector. According to this, the deviation
from maximality
of the solar mixing may be related to the quark mixing angle
$\theta_C\equiv \theta^{q}_{12}$ and is based on the
observation that the mixing angle responsible
for solar neutrino oscillations, $\theta_{\odot}\equiv \theta^\nu_{12}$
satisfies an interesting
complementarity relation with the corresponding angle in the quark sector
$\theta_{Cabibbo}\equiv \theta^q_{12}$ i.e. $\theta^\nu_{12}+\theta^q_{12}
\simeq \pi/4$.
 While it
is quite possible that this relation is purely accidental or due to some
other dynamical effects, it is interesting
to pursue the possibility that there is a deep meaning behind it
and see where it leads. It has been shown in a recent paper that if
Nature is quark lepton unified at high scale, then a relation between
$\theta^\nu_{12}$ and $\theta^q_{12}$ can be obtained in a natural manner
provided the neutrinos obey the inverse hierarchy\cite{fram}. It predicts
$sin^2\theta_\odot\simeq 0.34$ which agrees with present data at the
2$\sigma$ level. It also predicts a large $\theta_{13}\sim 0.18$, both of
which are predictions that can be tested experimentally in the near
future.

\subsection{Large mixing for Degenerate neutrinos:}
In this case, there are two ways to proceed: one may add the unit matrix
to either of the above mass matrices to understand large mixings or look
for some dynamical ways by which large mixings can arise. It
turns that in this case, one can generate large
mixings out of small mixings\cite{babu1,balaji} purely as a consequence
of radiative corrections. We will call this possibility radiative
magnification.

Let us illustrate the basic mechanism for the case of two generations.
The mass matrix in the $\nu_\mu-\nu_\tau$
sector\cite{balaji} cab written in the flavor basis as:
\begin{eqnarray}
{M_F(M_R)} =  U(\theta)
       \left(\begin{array}{cc} m_1 & 0 \\ 0 & m_2 \end{array}
\right) U(\theta)^{\dagger}
\label{uudag}
\end{eqnarray}
where $U(\theta)~=  \left(\begin{array}{cc} C_\theta & S_\theta \\
-S_\theta &
C_\theta \end{array} \right)$.
 This mass matrix is defined at the
seesaw (GUT) scale, where we assume the mixing angles to be small. As we
extrapolate this mass matrix down to
the weak scale, radiative corrections modify it to the form\cite{babu1}
\begin{eqnarray}
\cal{M_F (M_Z)} ~=~ \cal{R}\cal{M_F (M_R)}\cal{R}
\label{mf-fin}
\end{eqnarray}
where $\cal{R}~=~ \left(\begin{array}{cc} 1+\delta_\mu & 0 \\ 0 &
1+\delta_\tau \end{array} \right)$. Note that $\delta_{\mu} \ll
\delta_\tau$. So if we ignore $\delta_\mu$, we find that the $\tau\tau$
entry of the $\cal{M_F(M_Z)}$ is changed compared to its value at the
seesaw scale. If the seesaw scale mass eigenvalues are sufficiently close
to each other, then the two eigenvalues of the neutrino mass matrix at the
$M_Z$ scale can be same leading to maximal mixing (much like MSW matter
resonance effect) regardless what the values of the mixing angles at the
seesaw scale are. Thus at the seesaw scale, lepton mixing angles can even
be same as the quark
mixing angles as a quark-lepton symmetric theory would require. We call
this phenomenon radiative magnification of mixing angles. It requires no
assumption other than the near degeneracy of neutrino mass eigenvalues and
that all neutrinos have same CP (or all mass terms have same sign).
This provides a new dynamical mechanism to understand large mixings.

This radiative magnification mechanism has recently been generalized to
the case of three
neutrinos\cite{parida}, where assuming the neutrino mixing angles at the
seesaw scale to be same as the quark mixing angles renormalization group
extrapolation alone leads to large solar and atmospheric as well as small
$\theta_{13}$ at the weak scale provided the common mass of the neutrinos
$m_0\geq 0.1$ eV. The values of the mixing angles at the weak scale are
in agreement with observations i.e.  while both the solar and atmospheric
mixing angles become
large, the $\theta_{13}$ parameter remains small ($0.08$).

\begin{figure}[h]
\begin{center}
\includegraphics{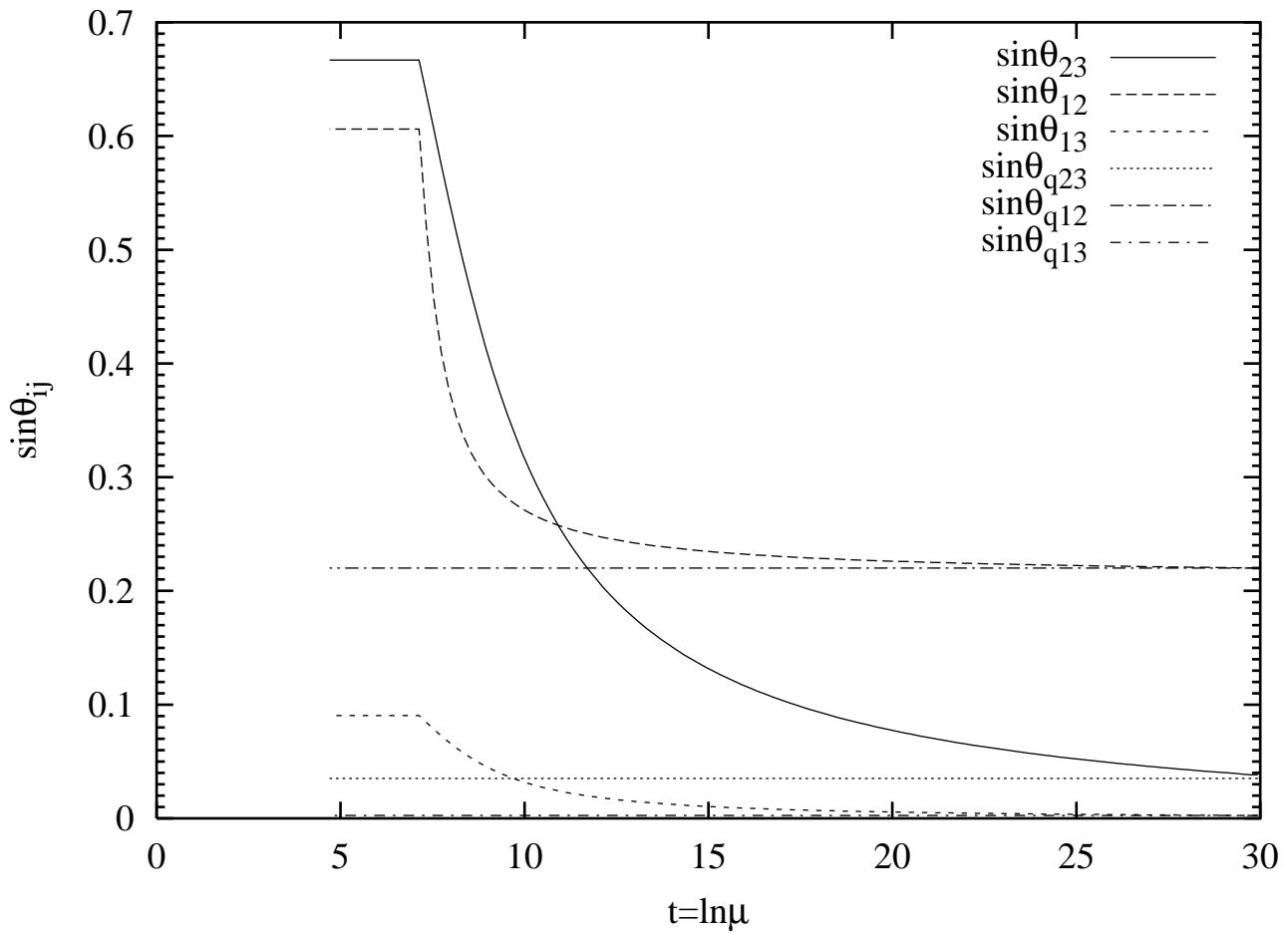}
\caption{High scale mixing unification and Radiative magnification
of mixing angles for degenerate neutrinos. Note that while the lepton
mixing angles get magnified, the quark mixings do not due essentially to
the hierarchical pattern of masses.}
\end{center}
\end{figure}

As already noted, an important
prediction of this model is that the common mass of the neutrinos must be
bigger than 0.1 eV, a prediction that can be tested in the proposed
neutrinoless double beta decay experiments.

There are many other proposals to understand large neutrino mixings; see
for instance \cite{anarchy} as one class of models and others
summarized in \cite{king}. An important physical insight one gains from
the various ways (models) of ensuring large $\theta_A$ and large
$\theta_{\odot}$ is that each have their characetristic predictions for
$\theta_{13}$ as well as deviation from solar as well as atmospheric
neutrino mixing from maximality. For a sample of these predictions, see
\cite{deviation}. As further high precision neutrino experiments are
carried out, they can be used to test the various ideas hopefully leading
to new insight into the nature of new physics.

\section{SEESAW MECHANISM AND GRAND UNIFICATION}
A naive estimate of the seesaw scale (or the scale of B-L symmetry) can be
obtained by using the $\Delta m^2_A\sim 2.5\times 10^{-3}$ eV$^2$ and the
seesaw formula $m_3\simeq\simeq \sqrt{\Delta m^2_A}\simeq
\frac{m^2_{33,D}}{M_R}$. The value of
$m^2_{33,D}$ is of course unknown however in the context of specific
models that unifiy quarks and leptons, one expects this to be of order 100
GeV  or so. Using this, one can conclude that $M_R\simeq 10^{14}-10^{15}$
GeV. This value is tantalizingly close to the scale of coupling
unification in supersymmetric theories, which is around $10^{16}$
GeV\cite{unif}. A natural possibility is therefore to discuss the seesaw
mechnism within the framework of grand unified theories.

As a simple possibility, one may consider the supersymmetric grand unified
theories. In this class of models, one assigns matter and Higgs to the
representations as follows: matter per generation are assigned to
$\bar{5}\equiv \bar{F}$ and $10\equiv 10$ dimensional representations
whereas the Higgs fields are assigned to $\Phi\equiv 45$, $H\equiv {5}$
and
$\bar{H}\equiv \bar{5}$ representations.

\noindent{\it \underline{Matter Superfields:}}
\begin{equation}
\bar{F} =\left(\begin{array}{c}
d^c_1\\ d^c_2\\ d^c_3 \\e^-\\ \nu\end{array}\right)\\ \nonumber
  ; T \{ 10 \}=\left(\begin{array}{ccccc}
0 & u^c_3 & -u^c_2 & u_1 & d_1\\
-u^c_3 & 0 & u^c_1 &u_2 & d_2 \\
u^c_2 & -u^c_1 & 0 & u_3 & d_3 \\
-u_1 & -u_2 & u_3 & 0 & e^+\\
-d_1 & -d_2 & -d_3 & -e^+ & 0 \end{array} \right)
\end{equation}
In the following discussion, we will choose the group indices as
$\alpha, \beta$ for SU(5);
(e.g.$ H^\alpha, \bar{H}_\alpha, \bar{F}_{\alpha}
T^{\alpha\beta}= -T^{\beta\alpha}$ );
$i,j,k..$ will be used for $ SU(3)_c $ indices and
$p,q$ for $ SU(2)_L$ indices.

To discuss symmetry breaking and other dynamical aspects of the model, we
choose the superpotential to be:
\begin{equation}
W = W_Y + W_G + W_h + W\prime
\end{equation}
where
\begin{eqnarray}
W_Y = h_u^{ab} \epsilon_{\alpha\beta\gamma\delta\sigma} T_a^\alpha\beta
T_b^{\gamma\delta} H^\sigma + h_d^ab T^{\alpha\beta} \bar{F}_\alpha
\bar{H}_\beta
\end{eqnarray}
($a,b$ are generation indices). This part of the superpotential is
resposible for giving mass to the fermions.
Effective superpotential for matter sector at low energies then looks
like:
\begin{eqnarray}
W_{matter}~=~ h_u QH_uu^c + h_d QH_d d^c + h_l LH_d e^c +\mu H_u H_d
\end{eqnarray}
Note that $h_d$ and $h_l$ arise from the $T\bar{F}\bar{H}$ coupling
and this satisfy the relation $h_d=h_l$. This relation leads to mass
equalities at the GUT scale of the form: $m_e=m_d$; $m_\mu = m_s$ and
$m_\tau~=~m_b$. These relations have to be extrapolated to the weak scale
to compared with observations. While the extrapolation for the third
generation is in very good agreement with data, it is far from
observations for the first and the second generations. This is of course a
problem for minimal SUSY SU(5) GUT. This model of course has the problem
of R-parity breaking by dimension 4 operators, which can lead to very
rapid proton decay.

Ignoring the fermion mass and R-parity problems, we can proceed to see how
it
accomodates light neutrino masses. Again as in the case of standard model,
one can add three right handed neutrinos as singlets to the  SU(5) theory
and use the seesaw mechanism to generate small neutrino masses. The
problem however is that we have no reason to choose the mass scvale of the
RH neutrinos to be at the GUT scale. In fact a natural choice would be the
Planck scale. Thus while as a practical model for neutrino masses,
SU(5) GUT theory may be OK, it faces the naturalness problem with respect
to the seesaw scale.

\section{SO(10) GRAND UNIFICATION AND NEUTRINO MIXINGS:}
We will now consider the SO(10) model, which as wel will see has a number
of virtues that make it just the right GUT model for neutrino masses.

First point to note is that the
{\bf 16} dimensional spinor representation of SO(10)
consists of all fifteen
standard model fermions plus the right handed neutrino arranged according
to the it $SU(2)_L\times SU(2)_R\times SU(4)_c$ subgroup as
follows:
\begin{eqnarray}
{\bf \Psi}~=~\pmatrix{u_1 & u_2 & u_3 & \nu\cr d_1 & d_2 & d_3& e}
\end{eqnarray}
We can take three such spinors for three fermion families. The presence of
the Pati-Salam subgroup $SU(4)_c$ allows relations between the neutrino
couplings and the quark couplings thereby raising the possibility there
will be fewer parameters in the model and more predictivity in the
neutrino sector, compared to the simple seesaw formula.

Secondly, SO(10) contains the B-L symmetry as a gauge symmetry. since
the mass of the righthanded neutrino breaks B-L symmetry, it has to be
constrained from above by the GUT scale, thus eliminating the hierarchy
problem that emerged in the SUSY SU(5) case.

 In order to implement the seesaw mechanism, one must
break the B-L symmetry, since the right handed neutrino mass breaks this
symmetry. One implication of this is that the seesaw scale is at or
below the GUT scale. Secondly in the context of supersymmetric
SO(10) models, the way B-L breaks has profound consequences for
low energy physics. There are two ways to break B-L in SUSY
SO(10) models: (i) by {\bf 16} Higgs or (ii) by {\bf 126} Higgs. Below we
give a comparision between the two ways and the present recent results
that follow from the second way.

\subsection{Breaking B-L: {\bf 16} vrs {\bf 126}}
 If B-L is broken by a Higgs field
belonging to the {\bf 16} dimensional Higgs field (to be denoted by
$\Psi_H$), then the field that
acquires a nonzero vev has the quantum numbers of the $\nu_R$ field
i.e. B-L breaks by one unit.  If we recall the definition of R-parity
i.e. $R_p~=~(-1)^{3(B-L)+2S}$, we see that this vev hav has
$R_p~=~-1$. This implies that the effective MSSM below the GUT scale in
such theories will break R-parity. To see how dangerous these operators
can be, note that in this case higher dimensional operators
of the form $\Psi\Psi\Psi\Psi_H$ are the ones that lead to R-parity
violating operators in the effective low energy MSSM theory. They then
lead to operators such as $QLd^c, u^cd^cd^c$ etc. Together these two
 can lead to large breaking of lepton and baryon number symmetry  with
a strength of $\left(\frac{v_{B-L}}{M_PM_{\tilde{q}}}\right)^2$. They lead
to unacceptable rates for proton decay (e.g. $\tau_p \leq $ sec.). This
theory
also has no dark matter candidate without making additional assumptions.

Secondly, in this class of theories, the right handed neutrino mass is
assumed to arise out of operators of the form
$\lambda \Psi\Psi\Psi_H\Psi_H/M_P$.
To get $M_R$ of order $10^{14}$ GeV, we would need to assume $\lambda
\simeq 1$. However, it is well known that similar dimension 5 operators
$\lambda'\Psi\Psi\Psi\Psi/M_P$ can also lead tp proton decay rate in
contradiction with observations unless $\lambda'\leq 10^{-6}$. This raises
a naturalness question which is why some operators have coefficients of
order one whereas others have coefficients of order $1o^{-6}$.

On the other hand, if one break B-L by a {\bf 126} dimensional Higgs
field, none of these problems arise.To see this note that the member of
this {\bf 126} multiplet that acquires vev has $B-L=2$ and
 therefore it leaves R-parity as an automatic symmetry of the low energy
Lagrangian. There is a naturally stable dark matter in this
case. Secondly, in this case, all fermion masses (including the right
handed neutrinos) arise from dimension four operators e.g. $\psi \psi
\bar{\bf 126}$ gives rise to right handed neutrino masses. Therefore we
can safely put all dimension five operators to have couplings less than
$10^{-6}$ without any problem.

A further point is that, any theory with asymptotic
parity symmetry
leads to type II seesaw formula. It turns out that if the B-L symmetry is
broken by {\bf 16} Higgs fields, the first term in the type II seesaw
(effective triplet vev induced term) becomes very small compared to the
type I term. On the other hand, if B-L is broken by a {\bf 126} field,
then the first term in the type II seesaw formula is not necessarily small
and can in principle dominate in the seesaw formula. We will discuss a
model of this type below.

\subsection{Minimal SO(10) with a single {\bf 126} as a predictive model
for neutrinos}
The basic ingredients of this model are that one considers only two Higgs
multiplets that contribute to fermion masses i.e. one
{\bf 10} and
one {\bf 126}. A unique property of the {\bf 126}
multiplet is that it not only breaks the B-L symmetry and therefore
contributes to
right handed neutrino masses, but it also contributes to charged fermion
masses by virtue of the fact that it contains MSSM doublets which mix with
those from the {\bf 10} dimensional multiplets and survive down to the
MSSM scale. This leads to a tremendous reduction of  the number of
arbitrary parameters, as we will see below.

There are only two Yukawa coupling matrices in this model: (i) $h$ for
the {\bf 10} Higgs and (ii) $f$ for the {\bf 126} Higgs.
SO(10) has the property that the Yukawa couplings involving the {\bf 10}
and {\bf 126} Higgs representations are symmetric. Therefore
if we assume that CP violation arises from other sectors of the theory
(e.g. squark masses) and work in a basis where one of these two sets
of Yukawa coupling matrices is diagonal, then it will have
only nine parameters. Noting the fact that the (2,2,15) submultiplet of
{\bf 126} has a pair of standard model doublets that contributes to
charged fermion masses.
In SO(10) models of this type,
the {\bf 126} multiplet contains two parity partner Higgs submultiplets
(called $\Delta_{L,R}$) which couple to $\nu_L\nu_L$ and $N_RN_R$
respectively and after spontaneous symmetry breaking lead to the type II
seesaw formula for neutrinos, which plays an important role in magnifying
the neutrino mixings despite quark-lepton unification\cite{goran,goh}.

As we will see a further advantage of using {\bf 126} multiplet is that it
unifies the charged fermion Yukawa couplings with
the couplings that contribute to righthanded as well as lefthanded
neutrino masses, as long as we do not include nonrenormalizable
couplings in the superpotential. This can be seen as
follows\cite{babu}: it is the set {\bf 10}+${\bf
\overline{126}}$ out of which the MSSM Higgs doublets emerge; the
later also contains the multiplets $(3,1,10)+(1,3,\overline{10})$
which are responsible for not only lefthanded but also the right
handed neutrino masses in the type II seesaw formula.
Therefore all fermion masses in the model are arising
from only two sets of $3\times 3$ Yukawa matrices one denoting the
{\bf 10} coupling and the other denoting ${\bf \overline{126}}$
couplings.
The SO(10) invariant superpotential giving the Yukawa couplings of the
{\bf
16} dimensional matter spinor $\psi_i$ (where $i,j$ denote generations)
with the Higgs fields $H_{10}\equiv
{\bf 10}$ and $\Delta\equiv {\bf \overline{126}}$.
\begin{eqnarray}
{W}_Y &=&  h_{ij}\psi_i\psi_j H_{10} + f_{ij} \psi_i\psi_j\Delta
\end{eqnarray}
 In terms of the GUT scale Yukawa couplings, one can write the
fermion mass matrices (defined as ${ L}_m~=~\bar{\psi}_LM\psi_R$) at
the seesaw scale as:
\begin{eqnarray}
M_u~=~ h \kappa_u + f v_u \\\nonumber
M_d~=~ h \kappa_d + f v_d \\  \nonumber
M_\ell~=~ h \kappa_d -3 f v_d \\  \nonumber
M_{\nu_D}~=~ h \kappa_u -3 f v_u \\\nonumber
\end{eqnarray}
where $\kappa_{u,d}$ are the vev's of the up and down standard model
type Higgs fields in the {\bf 10} multiplet and $v_{u,d}$ are the
corresponding vevs for the same doublets in {\bf 126}.
Note that there are 13 parameters in the above equations and there are 13
inputs (six quark masses, three lepton masses and three quark mixing
angles and weak scale). Thus all parameters of the model that go into
fermion masses are determined.

These mass sumrules provide the first
important ingredient in discussing the neutrino sector.
 To see this let us note that they lead to the following sumrule
involving the
charged lepton, up and down quark masses:
\begin{equation}\label{main}
    k \tilde{M}_l=r \tilde{M}_d+\tilde{M}_u
\end{equation} where $k$ and $r$ are
functions of the symmetry breaking parameters of the model. It is clear
from the above equation that smallquark mixings imply that the
contribution the charged leptons to the neutrino mixing matrix
i.e. $U_{\ell}$ in the formula $U_{PMNS}~=~U^{\dagger}_{\ell}U_{\nu}$
is close to identity and the entire contribution therefore comes from
$U_\nu$. Below we show that $U_nu$ has the desired form with
$\theta_{12}$ and $\theta_{23}$ large and $\theta_{13}$ small.

\subsection{Maximal neutrino mixings from type II seesaw}
In order to see how the type II seesaw formula provides a simple way to
understand large neutrino mixings in this model, note that in certain
domains of the parameter space of the model, the second matrix in the type
II seesaw formula can much smaller than the first term. This can happen
for instance when $V_{B-L}$ scale is much higher than $10^{16}$ GeV. When
this happens, one can derive the sumrule
\begin{eqnarray}
{ M}_{\nu} &=& a(M_{\ell}-M_d)
\label{key}\end{eqnarray}
This equation is key to our discussion of the neutrino masses and mixings.

Using Eq. (\ref{key}) in second and third generation sector, one
can understand how large mixing angle emerges.

Let us first consider the two generation case \cite{goran}. The known
hierarchical
structure of quark and lepton
masses as well as the known small mixings for quarks suggest that
the matrices $M_{\ell,d}$ for the second and third generation have

\begin{eqnarray}
M_{\ell}~\approx~m_\tau\left(\begin{array}{cc}\lambda^2
&\lambda^2\\
\lambda^2 & 1\end{array}\right)\\ \nonumber
M_q ~\approx~m_b \left(\begin{array}{cc}\lambda^2 & \lambda^2\cr \lambda^2
& 1\end{array}\right)
\end{eqnarray}
where $\lambda \sim 0.22$ (the Cabibbo angle).
 It is well known that in supersymmetric
theories, when low energy quark and lepton masses are extrapolated
to the GUT scale, one gets approximately that $m_b\simeq m_\tau$.
One then sees from the above sumrule for neutrino masses Eq.
(\ref{key}) that there is a cancellation in the $(33)$ entry of the
neutrinomass matrix and all entries are of
same order $\lambda^2$ leading very naturally to the atmospheric
mixing angle to be large. Thus one has a natural understanding of
the large atmospheric neutrino mixing angle. No extra symmetries
are assumed for this purpose.

For this model to be a viable one for three generations, one must
show that the same $b-\tau$ mass convergence at GUT scale also
explains the large solar angle $\theta_{12}$ and a small
$\theta_{13}$. This has been demonstrated in a recent
paper\cite{goh}.

To see how this comes about, note
that in the basis where the down
quark mass matrix is diagonal, all the quark mixing effects are
then in the up quark mass matrix i.e. $M_u ~=~ U^T_{CKM}M^d_u
U_{CKM}$. Using the Wolfenstein parametrization for quark mixings,
we can conclude that that we have
\begin{eqnarray}
M_{d}~\approx ~m_{b}\left(\begin{array}{ccc}\lambda^4 & \lambda^5
&\lambda^3\\ \lambda^5 & \lambda^2& \lambda^2 \\ \lambda^3 & \lambda^2 &
1\end{array}\right)
\end{eqnarray}
and $M_{\ell}$ and $M_d$ have roughly similar pattern due to the
sum rule . In the above equation, the matrix elements are
supposed to give only the approximate order of magnitude. As we
extrapolate the quark masses to the GUT scale, due to the fact
just noted i.e. $m_b-m_\tau \approx m_{\tau}\lambda^2$,
 the neutrino mass matrix $M_\nu~=~c(M_d-M_\ell)$
takes roughly the form:
 \begin{eqnarray}
M_{\nu}~=~c(M_d-M_\ell)\approx ~m_0\left(\begin{array}{ccc}\lambda^4 &
\lambda^5
&\lambda^3\\ \lambda^5 & \lambda^2 & \lambda^2 \\ \lambda^3 & \lambda^2
& \lambda^2\end{array}\right)
\end{eqnarray}
 It is then easy to see from this mass matrix that both the $\theta_{12}$
(solar angle) and $\theta_{23}$ (the atmospheric angle) are
large. It also turns out that the ratio of masses $m_2/m_3\approx \lambda$
which explains the milder hierarchy among neutrinos compared to
that among quarks. Furthermore, $\theta_{13}\sim \lambda$.
A detailed numerical analysis for this modelhas been carried out in
\cite{goh} and it substantiates the above analytical reasoning and
makes detailed predictions for the mixing angles\cite{goh}. We find that
the predictions for $sin^22\theta_{\odot}\simeq 0.9-0.94$,
$sin^22\theta_A \leq 0.92$, $\theta_{13}\sim 0.16$ and $\Delta
m^2_{\odot}/\Delta m^2_A\simeq 0.025-0.05$ are all in agreement with
data. Furthermore the prediction for $\theta_{13}$ is in a range that can
be tested partly in the MINOS experiment but more completely in the
proposed long baseline experiments.

\begin{figure}[h]
\begin{center}
\includegraphics{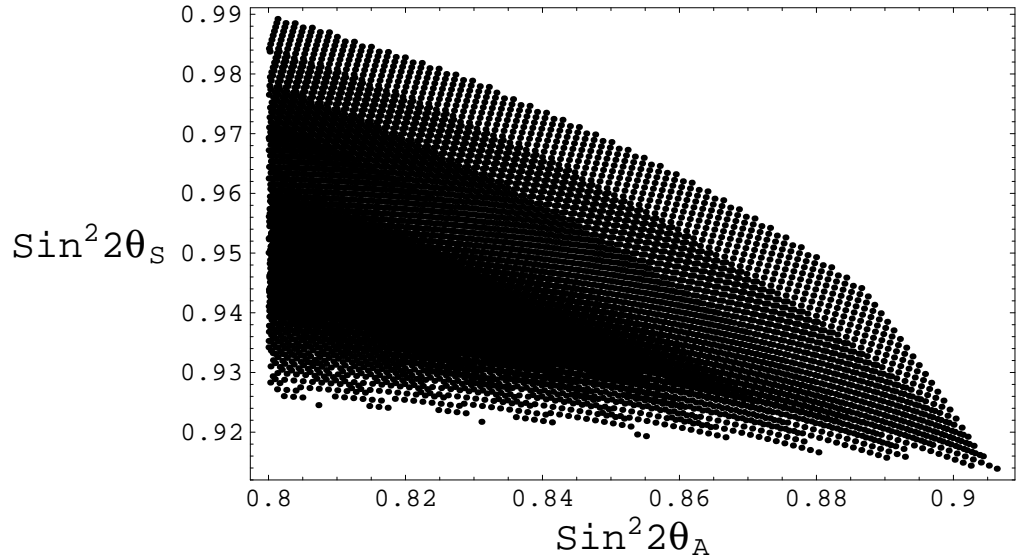}
\caption{$sin^22\theta_{12}$ ~vrs $sin^22\theta_{23}$; scatter
corresponds to different allowed quark mass values.}
\end{center}
\end{figure}

\begin{figure}[h]
\begin{center}
\includegraphics{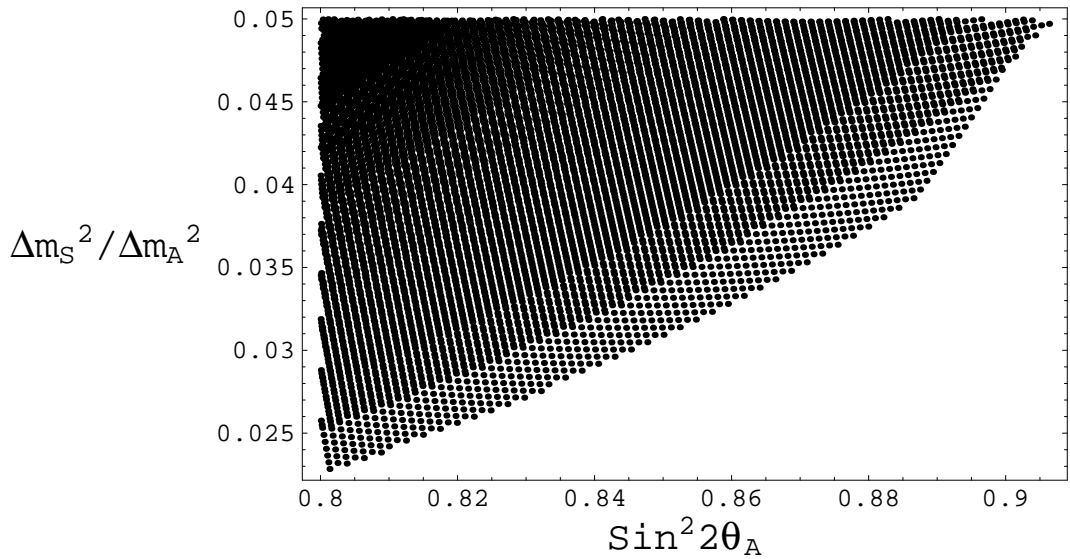}
\caption{scatter corresponds to uncertainty in quark mass values.}
\end{center}
\end{figure}

\begin{figure}[h]
\begin{center}
\includegraphics{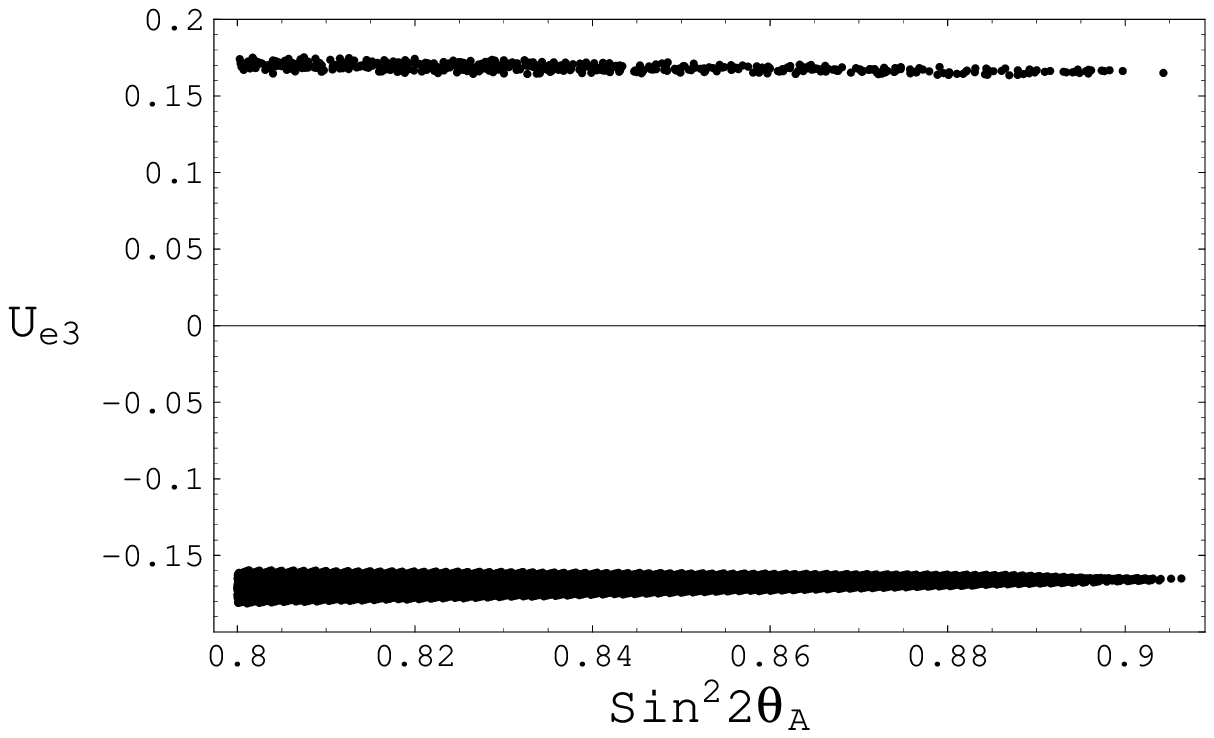}
\caption {$U_{e3}\equiv \theta_{13}$ and just below the present
upper limit: ``high'' value due to  no $\mu\leftrightarrow\tau$
symmetry (see before).}
\end{center}
\end{figure}

This model has been the subject of many further investigation including
such questions as to how to include CP violation, its predictions for
proton decay etc.\cite{other}.

We have not discussed the SO(10) models with {\bf 16} Higgs\cite{16} or
multiple {\bf 126} models\cite{chen}.

 \section{LSND and STERILE NEUTRINO}
The first need for sterile neutrinos came from attempts to
explain\cite{caldwell} apparent observations in the
Los Alamos Liquid Scintillation Detector (LSND) experiment\cite{lsnd} ,
of oscillations of $\bar{\nu}_\mu$'s from a stopped muon (DAR) as
well as of the $\nu_\mu$'s accompanying the muon in pion decay
(known as the decay in flight or DIF neutrinos) have apparently been
observed. The evidence from the
DAR is statistically
more significant and is an oscillation from $\bar{\nu}_\mu$ to
$\bar{\nu}_e$. The mass and mixing parameter range that fits data is:
\begin{eqnarray}
 \Delta m^2 \simeq 0.2 - 2 eV^2; sin^22\theta \simeq 0.003-0.03
\end{eqnarray}
There are points at higher masses specifically at 6 eV$^2$ which are
also allowed by the present LSND data for small mixings.
KARMEN experiment at the Rutherford laboratory has very strongly
constrained the allowed parameter range of the LSND
data\cite{karmen}. Currently the
Miniboone experiment at Fermilab is under way to probe the LSND parameter
region\cite{louis} using $\nu_\mu$ beam.

Since the  $\Delta m^2_{LSND}$ is so different from that  $\Delta
m^2_{\odot, A}$, the simplest way to explain these results is to
add one\cite{caldwell} or two\cite{sorel} sterile neutrinos. For the case
of one extra sterile neutrino, there are two scenarios: (i) 2+2 and
(ii) 3+1. In the first case, solar neutrino oscillation is supposed to be
from $\nu_e$ to $\nu_s$. This is ruled out by SNO neutral current data. In
the second case, one needs a two step process where $\nu_{\mu}$
undergoes indirect oscillation to $\nu_e$ due to a combined effect of
$\nu_\mu-\nu_s$ and $\nu_e-\nu_s$ mixings (denoted by $U_{\mu,s}$ and
$U_{e s}$ respectively, rather than direct
$\nu_\mu-\nu_e$ mixing. As a result, the effective mixing angle in LSND
for the 3+1 case is
given by $4U^2_{e s}U^2_{\mu s}$ and the measured mass difference is given
by that between
 $\nu_{\mu,e}-\nu_s$ rather than  $\nu_\mu-\nu_e$. This scenario is
constrained by the fact that sterile neutrino
mixings are constrained by two sets of observations: one from the
accelerator searches for $nu_\mu$ and $\nu_e$ disappearance
and the second from big bang nucleosynthesis.

The bounds on $U_{es}$ and $U_{\mu s}$ from accelerator experiments
such as Bugey, CCFR and CDHS are of course dependent on particular
value of $\Delta m^2_{\alpha s}$ but for a rough order of magnitude, we
have $U^2_{es}\leq 0.04$ for $\Delta m^2 \geq 0.1$ eV$^2$ and $U^2_{\mu s}
\leq 0.2$ for $\Delta m^2 \geq 0.4$ eV$^2$\cite{bilenky}.

It is worth pointing out that SNO neutral current data has ruled out pure
$\nu_e-\nu_s$ transition as an explanation of solar neutrino puzzle by
8$\sigma$'s; however, it still allows as much as 40\% admixture of sterile
neutrinos and as we will see below, the sterile neutrinos could very well
be present at a subdominant level. Thus the 2+2 scenario seems to be
highly disfavored by observations, whereas the 3+1 scenario is barely
acceptable.

\subsection{Theoretical Implications of a confirmation of LSND}
If LSND results are confirmed by the Mini Boone experiment, it will
require substantial revision of our thinking about neutrinos. For one
thing one should expect deviation from the unitarity constraint on the
three active neutrino mixings. But a much more fundamental alteration in
our thinking about neutrinos may be called for.

One such interpretation is in terms of breakdown of CPT invariance resulting in
different spectra for neutrinos compared to
anti-neutrinos\cite{mura}. This hypothesis
is now pretty much in conflict with observations after the KamLand
experiment\cite{concha}.

Another possibility is that there may be one or more sterile neutrinos in
Nature. The immediate challenge for theory then is
 why a sterile neutrino which is a standard model singlet (since it does
not couple to the W and Z bosons) has a
mass which is so light. A priori one would expect it to be of order of the
Planck scale.

A model that very cleverly solves this problem is the mirror universe
model where it is postulated that coexisting with the standard model
particles and forces is an exact duplicate of it, the mirror sector to our
universe\cite{mirror}. The forces and matter in the mirror are different
but mirror
duplicates of what we are familiar with. We will not see the mirror
particles or forces because they do not couple to our forces or matter.
Gravity of course couples to both sectors.

In this models there will be analogs of $\nu_{e,\mu,\tau}$ in the mirror
sector ($\nu'_{e,\mu,\tau}$). They will play the role of the sterile
neutrinos. It is then clear that whatever mechanism keeps our neutrinos
light will keep the mirror neutrinos light too, thereby solving the most
vexing problem with sterile neutrinos. In Table II, we present the
particle assignment for the mirror model.

\begin{center}
\begin{tabular}{|c||c|}\hline
visible sector & mirror sector\\
{ $SU(3)_c\times SU(2)_L\times U(1)_Y$ }&{
$SU(3)'_c\times SU(2)'_L\times U(1)'_Y$ }\\ \hline { $W,
Z, \gamma, $ gluons} &{ $W', Z', \gamma', $ gluons'}\\
 {$\pmatrix{u_L\cr
d_L}$} & {
$\pmatrix{u'_L\cr d'_L}$}\\
{ $u_R, d_R$} & { $u'_R, d'_R$}\\
{$\pmatrix{\nu_L\cr e_L}$} &
{$\pmatrix{\nu'_L\cr e'_L}$}\\
{ $e_R, N_R$} & { $e'_R, N'_R$}\\\hline
\end{tabular}
\end{center}

The next question is how to understand why the sterile neutrinos required
to understand the LSND experiment so much heavier ($\sim $ eV) than the
acive neutrinos. This is explained in the mirror model by postulating that
the weak scale in the mirror sector is about 10 to 20 times heavier than
the familiar weak scale. We of course do not know the reason for this. But
it is possible that it is tied to another feature of mirror models
required for their viability i.e. asymmetric inflation which says that
the reheat temperature of the mirror sector after inflation is smaller
than the visible sector so that the number density of relativistic
particles at the epoch of inflation is very small and does not affect
the success of BBN. This is the so called asymmetric mirror
model\cite{bere}.

This model has all the ingredients needed to understand the LSND results.

\section{CONCLUSION}
Neutrino physics right now is at a crossroad. Enough important
discoveries have been made so that the knowledge of the masses and mixings
are playing a significant role in influencing the direction of new physics
beyond the standard model; on the other hand, to raise our knowledge about
neutrinos to the same level as quarks as well as to decide more precisely
the direction of new physics, we need more precise information about
masses and mixings than we currently have.

 For instance, on the theoretical side, the seesaw
mechanism for understanding the scale of neutrino masses is regarded as
the prime candidate not only due to its simplicity but also its
theoretical appeal. It is perhaps hinting at the
new physics beyond the standard model to be left-right symmetric (due to
the introduction of the right handed neutrinos) and possibly
also quark-lepton $SU(4)_c$ symmetric\cite{ps} as well as grand
unification. On the last point of grand inification, there are large
variety of models based on the simple SO(10) group. A generic prediction
of SO(10) models is the normal hierarchy or quasi-degeneracy for
neutrinos. So evidence for inverted hierarchy would be point strongly away
from the SO(10) route. Similarly, evidence for Dirac rather than Majorana
neutrino (see Table I) would be a strong blow to the simple seesaw
mechanism.

On the hand our understanding of mixing angles is far from
complete. No clear consensus seems to have emerged about any particular
idea. An exhaustive list of
scenarios have been suggested to understand the new and unusual pattern of
intra-family mixing among leptons with their characteristic predictions
for observable parameters such as $\theta_A-\pi/4$, $\theta_{13}$ and
$m_{eff}$ in $\beta\beta_{0\nu}$ decay. Experiments will play a crucial
role in clarifying the picture here. It is therefore important to
implement the proposals for measuring these observables in the next decade
and one will then not only
have a better understanding of the neutrinos but also a more definite
direction in the nature of new physics beyond the standard model.

Thus, there remain
enough important things about neutrinos that are unknown so that a healthy
investment in the field will definitely broaden the frontier of our
overall understanding of forces, matter and the Universe. For example, it
 is very likely going to throw light on such
important cosmological issues as the origin of matter and formation
of structure in the Universe.

Neutrino physics has been full of surprises and there may yet be some more
waiting. For instance, confirmation by Mini Boone of the LSND
results will be one such major branch point. While, our general discussion
of mixings will receive a small perturbation, the impact on the
theoretical side will be very major, raising a completely
new set of questions and opening a brand new frontier in particle physics.

Exciting times are ahead in neutrino physics !!

\begin{acknowledgments}

I would like thank the SSI organizers for the invitation to lecture, for
their support and for creating such a pleasant atmosphere for discussions.
This work is supported by the National Science Foundation grant
no. Phy-0354401.

\end{acknowledgments}

\end{document}